\documentclass{ujp} 
\usepackage[cp1251]{inputenc}
\usepackage[english,russian]{babel}

\nazva{EMISSION AND ABSORPTION OF LIGHT BY
HOT ELECTRONS IN MULTIVALLEY SEMICONDUCTORS (TERAHERTZ RANGE)}%

\udk{538 }

\nazvacol{EMISSION AND ABSORPTION OF LIGHT}%

\avtor{P.M. TOMCHUK, V.M. BONDAR}%
\avtorcol{P.M. TOMCHUK, V.M. BONDAR }%

\inst{Institute of Physics, Nat. Acad. of Sci. of Ukraine}%
\adr{(46, Nauky Ave., Kyiv 03680, Ukraine; e-mail:
ptomchuk@iop.kiev.ua) }

\begin{document}           
\setcounter{page}{666}%
\maketitl                 

\selectlanguage{english}
\anot{%
The angular dependences of the spontaneous emission by hot electrons
in multivalley semiconductors are studied theoretically and
experimentally using $n$-Ge as an example. We demonstrate that the
change in the scattering mechanism caused by the growth of electron
temperature can affect the angular scattering  dependence. In the
case when the heating field is applied along the symmetry axis of
the crystal [for $n$-Ge it is the axis (1,0,0,)], the angular
dependence of the emission was observed experimentally for the first
time, and the corresponding theory is proposed. When electrons have
identical concentration and temperature in every valley, the angular
dependence of emission is shown to be related to the violation of
symmetry of the energy distribution of electrons (from the
theoretical viewpoint, this effect means going beyond the scope of
the traditional diffusion approximation).}%

\section{Introduction}

Free charge carriers in semiconductors can both absorb and emit
light. The absorption prevails when a semiconductor is irradiated
with an external electromagnetic flux. The light emission dominates
when the semiconductor contains hot carriers and there is no
external irradiation. In the thermodynamic equilibrium, the energy
absorbed by the carriers in a unit time is equal to the energy
emitted by them. These processes are well studied, and their main
regularities are described in available monographies. Nevertheless,
in the multivalley semiconductors of $n$-Ge\ and\ $n$-Si types, the
absorption and emission by free electrons have some specific
features which have not been ultimately clarified yet. These
features are due to strong anisotropy of the electron energy
dispersion law, filling of several equivalent minima (valleys) in
the conduction band, and scattering anisotropy. The character of
scattering governs the mechanisms of light absorption and emission
because a \textquotedblleft third\textquotedblright\ body (besides
an electron and a photon) is required for the energy and momentum
conservation laws to be simultaneously valid in these processes. The
role of such a third body can be played by phonons, impurities, or
defects.

The peculiarities of light absorption and emission by electrons in
multivalley semiconductors were examined earlier in our works
\cite{1,2}.  The role of anisotropic scattering of electrons by
acoustic vibrations of the lattice was studied in more detail in
\cite{1}, while the role of impurity scattering by charged centers
was investigated in \cite{2}. We showed that there can be some
conditions when the polarization effects can be observed in the
phenomena of light absorption and emission, despite the cubic
symmetry of the multivalley semiconductors of $n$-Ge and $n$-Si
types. These polarization effects were studied theoretically in
\cite{1,2} and experimentally in \cite{1,3}. Nevertheless, some
features of polarization dependences remained unclear. In
particular, experiment demonstrates (see below) that the coefficient
which characterizes the polarization  dependence of emission can
change its sign as the temperature of hot electrons increases.
Another feature of emission which is to be explained is that, in
rather strong electric fields, polarization effects take place even
when the carrier-heating field is directed symmetrically with
respect to the valleys. Therefore, these issues require further
investigation.

The light absorption and emission by free carriers was studied
theoretically by various methods. As to the anisotropic scattering,
we presume that our method \cite{1,2} is most convenient and
effective. It is based on the collision integral and takes into
account the influence of electromagnetic waves on the electron
scattering.

The method has the following advantage. The quantum-mechanical case
(when the electron energy is lower than the light quantum energy)
and the classical case can be considered on unique grounds.
Moreover, the emission of free carriers induced by the wave field
can easily be found in addition to the absorption. Hence, one can
easily determine the spontaneous emission of hot electrons. Finally,
our approach allows a general expression for the emission
(absorption) to be derived in the form of integrals over the momenta
of all electrons and phonons (impurities), so that the order of
integration over the angles can be arbitrarily changed and
calculations can be simplified greatly in the case of anisotropic
scattering.

\section{Formulation of the Problem. Acoustic Scattering}

In this section, we start from the integral of collisions between
electrons and acoustic vibrations of the lattice, which already takes into
account the influence of the electromagnetic wave on the collision event. It
is known that, in semiconductors of the $n$-Ge\ and\ $n$-Si types, the
dispersion law for electron energy $\varepsilon _{\vec{p}}$ looks like
\begin{equation}
\label{eq1} \varepsilon _{\vec {p}} = {\sum\limits_{j = 1}^{3}
{{\frac{{p_{j}^{2} }}{{2m_{j}}} }}}  = {\frac{{p_{x}^{2} +
p_{y}^{2}}} {{2m_{ \bot}} } } + {\frac{{p_{z}^{2}}}
{{2m_{\parallel}} } },
\end{equation}
where $p_{j}$ are the components of the momentum, and $m_{\bot }$ and $%
m_{\parallel}$ are the transverse and the longitudinal effective
masses of an electron, respectively. To obtain the required collision
integral, it is necessary -- while deriving the kinetic equation
-- to use the wave functions $\Psi _{\vec{p}}$ of an electron in
the field of an electromagnetic wave rather than the basic wave
functions of a free electron:
\[\Psi _{\vec {p}} = {\frac{{1}}{{\sqrt {V}}} }\exp \left(
{{\frac{{i}}{{\hbar }}}\vec {p}\,\vec {r}} \right)\times
\]
\begin{equation}
\label{eq2} \times\exp {\left\{ { - {\frac{{i}}{{\hbar
}}}{\int\limits_{0}^{t} {d{t}'{\sum\limits_{j = 1}^{3}
{{\frac{{1}}{{2m_{j} }}}\left( {p_{j} - {\frac{{e_{0}}}
{{c}}}A_{j} ({t}')} \right)^{2}}}} }} \right\}}.
\end{equation}
Here, $V$ is the volume, $e_{0}$ the charge, $c$ the speed of light, $\vec{A}
$ the vector-potential of the electromagnetic wave which is adopted in the
form
\begin{equation}
\vec{A}=\vec{A}^{(0)}\,\cos \omega \,t , \label{eq3}
\end{equation}%
$\omega $ the electromagnetic wave frequency, and $t$ the time. Details on
the procedure of derivation of the collision integral for the case of the
impurity scattering, provided the dispersion law (\ref{eq1}) and taking the
influence of light on the collision event into account, can be found, e.g.,
in work \cite{2}. For the first time, such a collision integral for the
standard dispersion law was obtained in work \cite{4}.

Hence, the electron-phonon collision integral, which makes allowance for the
influence of an electromagnetic wave on a scattering event, looks like
\[
 If = {\sum\limits_{s = 1}^{3} {\,{\sum\limits_{\vec {q}} {}
}{\sum\limits_{l = - \infty} ^{\infty}  {W^{s}}}} } (\vec
{q})J_{l}^{2} \left( {{\frac{{e_{0} \gamma}} {{m_{ \bot}  \omega
c}}}} \right)\times\]
\[\times\Bigl\{{\left[ {f(\vec {p} + \hbar \vec {q}\,)(N_{\vec {q}}^{(s)} + 1) -
f(\vec {p}\,)N_{q}^{(s)}}  \right]}\times \]
 \[ \times \delta \,\left[ \varepsilon _{\vec {p} + \hbar \vec {q}} -
 \varepsilon_{\vec {p}} - \hbar \omega _{\vec {q}}^{(s)} - l\hbar \omega
\right] + \]
\[ + \left[ f(\vec{p}) - \hbar \vec {q}\,)N_{\vec {q}}^{(s)} - f(\vec {p}\,)(N_{\vec
{q}}^{(s)} + 1) \right] \times\]
\begin{equation}
\label{eq4} \times\delta \left[ \varepsilon _{\vec {p} - \hbar
\vec {q}} - \varepsilon _{\vec {p}} + \hbar \omega _{\vec
{q}}^{(s)} - l\hbar \omega   \right]\Bigr\}.
 \end{equation}

Here, $f(\vec{p})$ is the distribution function of electrons over the
momentum $\vec{p}$, $N_{\vec{q}}^{(s)}$ the distribution function of phonons
belonging to the $s$-th branch over the momentum $\hbar\vec{q}$, and $%
\,J_{l}\left( {{\frac{{e_{0}\gamma}}{{m_{\bot}\omega\,c}}}}\right) $ the
Bessel function of the $l$-order, which takes the influence of the wave on
the collision event into account. In the argument of the Bessel function,
the quantity
\begin{equation}
\gamma=\vec{A}^{(0)}\vec{q}+\left( {{\frac{{m_{\,||}}}{{m_{\bot}}}}-1}%
\right) \left( {\vec{A}^{(0)}\vec{l}_{0}}\right) \left( {\vec{q}\,\vec {l}_{0}}%
\right) ,  \label{eq5}
\end{equation}
where $\vec{l}_{0}$ is the unit vector which fixes the orientation of the
rotation axis of the effective-mass ellipsoid.

In Eq.~(\ref{eq4}), $W^{(s)}$ is the probability for an electron to be
scattered by an acoustic phonon of the $s$-th branch. The explicit form of
this quantity will be discussed below.

Leaving very low temperatures beyond the scope of the present consideration, we
may assume that the acoustic quantum energy $\hbar\omega_{\vec{q}%
}^{(s)}\ll T$, where the temperature of the lattice $T$ is
expressed in power units. Then,
\[
N_{\vec {q}}^{(s)} \approx T/(\hbar \omega _{\vec {q}}^{(s)})
 \gg 1.\]
In Eq.~(\ref{eq4}), in all the terms but the summand, where the
number of light quanta is equal to zero ($l=0$), the energy of acoustic quanta can be put equal to zero. Then, making the
substitution $\vec{q}\rightarrow -\vec{q}
$ in the second term of Eq.~(\ref{eq4}) and taking into account that $%
W^{(s)}(\vec{q}\,)=W^{(s)}(-\vec{q}\,)$, we obtain from this
equation that
\[
 If = I^{(0)}f + \sum\limits_{\vec {q}} \sum\limits_{s = 1}^{3}
\frac{2T}{\hbar \omega _{\vec {q}}^{(s)}}W^{(s)}(\vec
{q}\,)\times\]
\[ \times{\sum\limits_{l = - \infty} ^{\infty}}\!\!{}^{\prime} J_{l}^{2}
(\frac{e_{0} \gamma}{m_{ \bot}  \omega c})\,\left[ f(\vec {p} +
\hbar \vec {q}\,) - f(\vec {p}\,) \right]\times \]
\begin{equation}
\label{eq6}
 \times \delta \,\left[\varepsilon _{\vec {p}
+ \hbar \vec {q}} - \varepsilon _{\vec {p}} - l\hbar \omega
\right].
\end{equation}
In this expression, $I^{(0)}$ denotes the value of the integral $I$ in Eq.~%
\ref{eq4}) at $l=0$. This term makes allowance for the inelasticity of
acoustic scattering. The prime in the sum over $l$ means that the term with $%
l=0$ is omitted.

Changing in Eq.~(\ref{eq6}) from the sum over the index $\vec{q}$
to the integral over the variable
${\vec{p}}^{\,\prime}=\vec{p}+\hbar\vec{q}$ brings us to the
result
\[
 If = I^{(0)}f + \int {d{\vec {p}}\,'} {\sum\limits_{l = - \infty} ^{\infty}
{}} \!\!{}^{\prime}\, W_{a} (\vec {q}\,)\,J_{l}^{2} ({\frac{{e_{0}
\gamma}} {{m_{ \bot } \omega \,c}}})\,\times \]
\begin{equation}
\label{eq7}
 \times{\left[ {f({\vec {p}}\,') - f(\vec
{p})} \right]} \delta {\left[ {\varepsilon _{{\vec {p}}'} -
\varepsilon _{\vec {p}} - l\hbar \omega} \right]}.
\end{equation}
Here, we introduced the notation
\begin{equation}
\label{eq8} ^{}W_{a} (\vec {q}\,)\, = {\frac{{V}}{{(2\pi \hbar
)^{3}}}}{\sum\limits_{s = 1}^{3} {{\frac{{2T}}{{\hbar \omega
_{\vec {q}}^{(s)}}} }}} W^{(s)}(\vec {q}\,).
\end{equation}
The explicit expression for the probability of the electron scattering
by all three branches of acoustic vibrations can be obtained
in the deformation potential approximation \cite[page 162]{5}. The
calculation gives
\[W_{a} (\vec {q}\,) = \frac{T}{4\pi ^{2}\hbar ^{4}\rho}
\left\{ \frac{1}{s_{\vert \,\vert} ^{2}} \right.\left[
\sum\nolimits_{d}   + \sum\nolimits_{u}  \left( \frac{\vec {l}_{0}
\vec {q}}{q} \right)^{2} \right]^{2} +\]
\begin{equation}
\label{eq9} + \frac{\sum\nolimits_{u}^{2}}{s_{ \bot} ^{2}}\left[ 1
- \left(\frac{\vec {l}_{0} \vec {q}}{q} \right)^{2} \right]\left.
\left( \frac{\vec {l}_{0} \vec {q}}{\vec {q}} \right)^{2}
\right\}.
\end{equation}
In this expression, $\rho $ is the density; $s_{\parallel }$ and
$s_{\bot }$ are
the speeds of longitudinal and transverse sound, respectively; $\Sigma _{d}${%
\ and }$\Sigma _{u}$ are the constants of deformation potential; and $\vec{l}%
_{0}$ is the unit vector that fixes the valley orientation (the rotation
axis of the mass ellipsoid).

Now, let us write down the expression for the energy exchanged by electrons and
acoustic phonons in a unit time in the presence of an
electromagnetic wave:
\[
 P = \int {d\vec {p}\,\varepsilon_{\vec{p}}\,\hat {I}\,f(\vec {p}\,)} = P_{0} +
  \hbar \omega \sum\limits_{l = - \infty} ^{\infty}  l\int d\vec {p}\,f
 (\vec{p}\,)\times \]
\begin{equation}
\label{eq10} \times\int d{\vec {p}\,}'\, W_{a}  (\vec
{q}\,)\,J_{l}^{2} \left( \frac{e_{0} \gamma}{m_{ \bot}  \omega
\,c} \right)\delta \left[ \varepsilon _{\vec {p}\,'} - \varepsilon
_{\vec {p}} - l\hbar \omega \right].
\end{equation}
Here, the term $P_{0}$ is associated with the first term in Eq.~(\ref{eq6})
and characterizes the energy that the electron gas (if it is hot) transfers
to acoustic vibrations of the lattice for a unit time.

In what follows, we confine ourselves to the consideration of single-quantum
processes only, i.e., we consider only those terms in sum (\ref{eq10})
which correspond to the values $l=\pm 1$ only. Moreover, if extremely strong
fields are not examined, the argument of the Bessel function is considerably
smaller than unity, and, consequently, we may confined the consideration to
the first term of the expansion. Taking all those approximations
into account, we rewrite Eq.~(\ref{eq10}) as follows:
\[ P \approx P_{0} \pm \hbar \omega \int d\vec {p}\,f(\vec
{p}\,)\times\]
\begin{equation}
\label{eq11} \times\int d{\vec {p}\,}'W_{a}   (\vec {q}\,)\,\left(
\frac{{e_{0} \gamma}} {{2m_{ \bot} \omega \,c}} \right)^{2}\delta
\left[ \varepsilon _{\vec {p}\,'} - \varepsilon _{\vec {p}} \mp
\hbar \omega  \right].
\end{equation}
The argument of the $\delta$-function in Eq.~(\ref{eq11}) demonstrates that
the energy of the electron after scattering is equal to $\varepsilon _{{\vec{%
p}}^{\prime }}=\varepsilon _{\vec{p}}\pm \hbar \omega $, i.e. expression (%
\ref{eq11}) describes both the single-quantum radiation absorption and
the single-quantum emission.

To move further, we should specify the momentum-distribution function of
electrons $f(\vec{p})$. Hence, we consider multivalley semiconductors of the
$n$-Ge\ and\ $n$-Si types, in which electrons occupy lower (equivalent)
valleys. The distribution function for the electrons in the $i$-th valley
is supposed to be Maxwellian-like, but with its own electron concentration $%
n_{i}$ and own electron temperature $T_{i}$:
\begin{equation}
\label{eq12} f^{(i)}(\vec {p}\,) = \frac{n_{i}}{(2\pi T_{i}
)^{3/2}\,m_{ \bot } \sqrt {m_{\parallel}} }\,e^{ - \varepsilon
_{\vec {p}}/ T_{i}}.
\end{equation}
The concentration $n_{i}$ and the temperature $T_{i}$ depend on a
number of external factors (pressure, electric field, irradiation)
and should be determined from the corresponding balance equations
for the concentration and the energy. In our theory, they are
considered as the known parameters. If the function $f(\vec{p})$ in
Eq.~(\ref{eq11}) denotes function (\ref{eq12}) and the quantity
$\gamma $ is determined from Eq.~(\ref{eq5}), then expression
(\ref{eq11}) describes the contribution of charge carriers from
the $i$-th valley to the processes of light absorption and
emission. Below, the quantities associated with the $i$-th valley
are designated by the subscript $i$. Substituting Eq.~(\ref{eq12})
into Eq.~(\ref{eq11}) and changing to the deformed coordinate
system, where the energy does not depend on angles (see details in
works \cite{1,2}), the integrals can be easily calculated, and, as
a result, we obtain
\[
 \Delta P_{i} ( + ) = \frac{2e_{0}^{2}}{3\sqrt {\pi}}\left\{
\frac{A_{\perp} ^{(0)2}}{m_{ \perp}  \tau _{ \perp} ^{(0)}}
 + \frac{{A_{\parallel} ^{(0)2} }}{{m_{\parallel} \tau
_{\parallel} ^{(0)}}} \right\}\frac{n_{i}
T_{i}^{3/2}}{T^{1/2}c^{2}\hbar \omega}\times \]
\begin{equation}
  \times \left\{ a_{i}^{3}
e^{a_{i}}\frac{{d}}{{da_{i}}}\left(\frac{{K_{ 1}
(a{}_{i})}}{{a_{i}}} \right) \right\}_{a_{i} = \hbar \omega/
T_{i}} ,
 \end{equation}
 \[
\Delta P_{i} ( - ) = - \exp \left( { - {\frac{{\hbar \omega}}
{{T_{i}}} }} \right)\Delta P_{i} ( + ).
\]
In expression (13), we took into account only the second term in Eq.~(%
\ref{eq11}), which describes single-quantum transitions (the term including $%
P_{0}$ is not of interest for us). The quantities $\Delta P_{i}(+)$ and $%
\Delta P_{i}(-)$ describe the contributions of the $i$-th valley to
absorption and emission, respectively.

In Eq.~(13), $K_{1}(a_{i})$ is the so-called Bessel function of an
imaginary argument, which has the asymptote
\begin{equation}
\label{eq13} K_{ 1} ^{} (a_{i} ) = {\left\{ {\begin{array}{l}
 {{\frac{{1}}{{a_{i}}} },\quad \quad\quad\quad\text{at}\quad a_{i} \to 0}, \\
 {\sqrt {{\frac{{\pi}} {{2a_{i}}} }} \;\,e^{ - a_{i}} ,\quad \text{at}\quad a_{i}
\to \infty} .  \\
 \end{array}} \right.}
\end{equation}
In addition, two new physical quantities, $\tau _{\parallel }^{(0)}$
and $\tau _{\bot }^{(0)}$, were introduced in Eq.~(13); these are
the so-called \textquotedblleft longitudinal\textquotedblright\
and \textquotedblleft transverse\textquotedblright , respectively,
components of the relaxation
time. Their expressions in terms of the parameters $m_{\Vert }$, $m_{\bot }$%
, $\Sigma _{d}$, and $\Sigma _{u}$ are rather cumbersome (see, e.g., \cite[%
page 166]{5}). Nevertheless, their relationships with the corresponding
components of the acoustic mobility tensor for cold (not heated) electrons are
simple, namely,
\begin{equation}
\label{eq14} \mu _{\parallel} ^{(a)} = {\frac{{4}}{{3\sqrt {\pi}}
} }\,{\frac{{e\tau _{\parallel} ^{(0)}}} {{m_{\parallel}} }
},\quad \quad \mu _{\perp} ^{(a)} = {\frac{{4}}{{3\sqrt {\pi}} }
}\,{\frac{{e\tau _{ \perp }^{(0)}}} {{m_{ \bot}} } }.
\end{equation}
In formulas (13), the quantity $A_{\parallel}^{(0)}$ means the
component of the $\vec{A}^{(0)}$ vector directed along the
rotation axis of the $i$-th mass ellipsoid, and $A_{\bot}^{(0)}$
the component perpendicular to this axis, i.e.
\begin{equation}
\label{eq15} (A_{\parallel} ^{(0)} )^{2} = (\vec {A}^{(0)}\vec
{l}_{0} )^{2}\;,\; (A_{\perp} ^{(0)} )^{2} = (A^{(0)})^{2} - (\vec
{A}^{(0)}\vec {l}_{0} )^{2}.
\end{equation}

In Introduction, we mention that, depending on the situation, the
process of absorption or emission of light by free charge
carriers can dominate. Therefore, before speaking about the polarization
dependences of emission by free carriers in multivalley semiconductors, let
us write down the absorption coefficient. It equals
\begin{equation}
\label{eq16} K = \frac{\sum\limits_{(i)} \{ {P_{i} ( + ) + P_{i} (
- )} \}}{\prod} = \frac{\sum\limits_{(i)} \left\{ 1 - e^{ - \hbar
\omega / T_{i}} \right\}P_{i} ( + )}{\prod}.
\end{equation}
Here, $\Pi $ is the electromagnetic flux that -- in the case of
absorption -- falls onto the semiconductor:
\begin{equation}
\label{eq17} \prod = \frac{\chi _{0}^{1/2}} {8\pi }\,\frac{\omega
^{2}}{c}\,(A^{(0)})^{2},
\end{equation}
$\chi _{0}$ being the static dielectric constant.

From Eqs.~(13) and (\ref{eq15})--(\ref{eq17}), we obtain
\[
 K = \frac{16\sqrt {\pi}} {3}\,\frac{e_{0}^{2}}{c\hbar
}\,\sum\limits_{i} \frac{n_{i} T_{i}^{3/2}} {\omega T^{1/2}}
\,\times\]
\[
\times\left\{ \frac{1}{m_{ \bot}  \tau _{ \bot} ^{(0)}} + \left[
\frac{1}{m_{\parallel}  \tau _{\parallel} ^{(0)}} - \frac{1}{m_{
\bot}  \tau _{ \bot} ^{(0)}} \right](\vec {g}_{0} \vec
{l}{}_{0})^{2} \right\}\times \]
\begin{equation}
\label{eq18}
 \times \left( {1 - e^{ - \hbar \omega / T{}_{i}}}
\right){\left\{ {a_{i} e^{a_{i}} {\frac{{d}}{{da_{i}}}
}\left({\frac{{K_{ 1} (a_{i}}} {{a_{i_{}}} } }\right)}
\right\}}_{a_{i} = \hbar \omega / 2T_{i}},
\end{equation}
where $\vec{g}_{0}$ is a unit vector which sets the polarization, i.e.
\[
\vec{A}^{(0)}=\vec{g}_{0}A^{(0)}.
\]

\section{Polarization Dependences of Spontaneous Emission by Hot
Electrons. Acoustic Scattering}

In this section, we discuss the spontaneous emission by
electrons. We consider the electron gas to be heated by means of
an electric field applied to the semiconductor. Therefore, both
the electron concentration $n_{i}$ in the $i$-th valley and the
corresponding electron temperature $T_{i}$ in it differ from the
relevant equilibrium values of those parameters. According to
formulas (13), the quantity $P_{i}(-)$ is responsible for the
emission of hot electrons from that valley induced by
the wave field. To derive the expression describing the spontaneous
emission, we will act in the following way.

First, we normalize the vector-potential of the wave, so that the volume $V$
should contain $N_{pb}$ photons, i.e. we assume that
\begin{equation}
\label{eq19} {\frac{{1}}{{V}}}N_{pb} \,\hbar \omega = {\frac{{\bar
{E}^{2}}}{{4\pi}} } \equiv {\frac{{1}}{{8\pi}} }\left(
{{\frac{{\omega}} {{c}}}} \right)^{2}\left( {A^{(0)}} \right)^{2}.
\end{equation}
Whence, we obtain
\begin{equation}
\label{eq20} A^{(0)} = 2c\left( {{\frac{{2\pi \hbar}} {{V\omega}}
}N_{pb}} \right)^{1/2}.
\end{equation}
We now substitute expression (\ref{eq20}) into Eq.~(13), put $%
N_{pb}=1 $ there, and multiply the expression obtained for $P_{i}(-)$ by the
final state density of the electromagnetic field in the unit interval of
frequencies and the solid angle $d\underline {o}$ to obtain
\begin{equation}
\label{eq21} d\rho (\omega ) = {\frac{{V}}{{\left( {2\pi c}
\right)^{3}}}}^{}\omega ^{2}\;d\underline {o}.
\end{equation}
As a result, the expression for the energy, which is radiated by electrons
from all valleys in a unit time, in a unit volume, and into the solid angle
$d\underline {o}$, looks like
\[
 W^{( - )} = - \frac{2e_{0}^{2}}{3\pi ^{5/2}c^{3}T^{1/2}}\sum\limits_{(i)}
n_{i}  T_{i}^{3/2} \Biggl\{ \frac{1}{m_{ \bot}  \tau _{ \bot}
^{(0)}} + \]
\[+\Biggl(\frac{1}{m_{\parallel}  \tau
_{\parallel} ^{(0)}} - \frac{1}{m_{ \bot}  \tau _{ \bot} ^{(0)}}
\Biggr)\,\left( {\vec {l}_{0} \vec {g}_{0}} \right)^{2}
\Biggr\}\times \]
\begin{equation}
\label{eq22}
 \times \left( a_{i}^{3}
\,e^{ - a_{i}} \frac{d}{da_{i}}\left( \frac{K_{ 1} (a_{i}
)}{a_{i}} \right) \right)_{a_{i} = \hbar \omega /2T_{i}}
\,d\underline {o}.
\end{equation}
Formula (\ref{eq22}) describes both the cases of classic (if\thinspace $%
a_{i}\ll 1$) and quantum-mechanical (if\thinspace $a_{i}\gg 1$)
emission. In particular, making use of asymptotics (\ref{eq13}),
Eq.~(\ref{eq22}) in the classical case yields
\[ W^{( - )} = \frac{4e_{0}^{2}}{3\,\pi
^{5/2}\,c^{3}\,T^{1/2}}\times\]
\begin{equation}
\label{eq23} \times\sum\limits_{(i)} {n\,_{i} \,} T_{i}^{3/2}
\left\{ \frac{\sin ^{2}\varphi _{i}}{m_{ \bot} \,\tau _{ \bot
}^{(0)}} + \frac{\cos ^{2}\varphi _{i}}{m_{\parallel} \,\tau
_{\parallel } ^{(0)}} \right\}\, d\,\underline {o}.
\end{equation}
Here, $\cos ^{2}\varphi _{i}=(\vec{l}_{0}\vec{g}_{0})^{2}$, so that $\sin
^{2}\varphi _{i}=1-(\vec{l}_{0}\vec{g}_{0})^{2}$.

Similarly, in the quantum-mechanical case, we have
\[ W^{( - )} = \frac{e_{0}^{2} (\hbar \omega )^{3}}{6\,\pi
^{2}\,c^{3}\,T^{1/2}}\times\]
\begin{equation}
\label{eq24} \times\sum\limits_{(i)} n\,_{i} e^{ - \hbar
\omega/T_{i}}\left\{ \frac{\sin ^{2}\varphi _{i}} {m_{ \bot}
\,\tau _{ \bot }^{(0)}} + \frac{\cos ^{2}\varphi _{i}}
{m_{\parallel} \,\tau _{\parallel} ^{(0)}} \right\}\,
\,d\,\underline {o}.
\end{equation}

The polarization dependence of the emission is characterized by the
angles $\varphi_{i}$ between the valley orientation (the rotation axis of
the mass ellipsoid) and the polarization of the emitted wave. To decipher these
angular dependences, it is necessary to find the distributions of the
parameters $n_{i}$ and $T_{i}$ over the valleys. Therefore, let us consider
a specific case where the electron-heating electric field is directed along
the (1,1,1)-axis.

As was mentioned earlier, isoenergetic surfaces around every minimum look like
ellipsoids of revolution (\ref{eq1}). The rotation axes of these ellipsoids
in $n$-Ge are given by the unit vectors
\[ \vec {l}_{1} = {\frac{{1}}{{\sqrt {3}}} }(1,1,1),\quad \vec
{l}_{2} = {\frac{{1}}{{\sqrt {3}}} }( - 1,1,1),\]
\begin{equation}
\label{eq25} \vec {l}_{3} = {\frac{{1}}{{\sqrt {3}}} }(1, - 1,1),
\quad \vec {l}_{4} = {\frac{{1}}{{\sqrt {3}}} }( - 1, - 1,1).
\end{equation}
Up to now, we used the notation $\vec{l}_{0}$ to designate the unit vector
which determined the rotation axis of an arbitrary $i$-th ellipsoid. Below,
we use the notation $\vec{l}_{k}$ $(k=1,2,3,4)$ to fix the direction
of specific valleys in $n$-Ge.

The energy supplied by a constant electric field $\vec{F}$ in a unit
time to the electrons of the $k$-th valley is equal to
\begin{equation}
\label{eq26} W_{\rm D} = n_{k} {\left\{ {\,\mu _{ \bot}  F^{2} +
(\mu _{\parallel}  - \mu _{ \bot}  )(\vec {l}_{k} \vec {F})^{2}}
\right\}}.
\end{equation}
Here, $\mu _{\parallel }$ is the longitudinal and $\mu _{\bot }$
the transverse component of the mobility tensor. From
Eq.~(\ref{eq26}), it is clear that, provided the field $\vec{F}$
is oriented along the (1,1,1)-direction, the energies released per
one electron in valleys 2, 3, and 4 are identical to each other,
being higher than that released in valley~1. So, the valley
represented by the unit vector $\vec{l}_{1}$ is \textquotedblleft
cold\textquotedblright , while the valleys characterized by the unit vectors $%
\vec{l}_{2}$, $\vec{l}_{3}$, and $\vec{l}_{4}$ are \textquotedblleft
hot\textquotedblright . In this situation, we have two sets of parameters: $%
(n_{1},T_{1})$ and $(n_{2}=n_{3}=n_{4},$ $T_{2}=T_{3}=T_{4})$.

But before analyzing the situation where the parameters for different
valleys can be different, let us consider the case with identical
concentrations and temperatures in all valleys. This situation takes place,
if the field is directed along the direction (1,0,0). In this case, if one
takes advantage of formula (\ref{eq23}) and, while carrying out summation
over $i$, takes unit vectors (\ref{eq25}) instead of the unit vector $\vec{l}%
_{0}$, the following formula is obtained:
\[
W^{( - )} \equiv W\,\,\left\{ {\,n_{e} \,,T_{e}}  \right\} =
\frac{16\;\,\,e_{0}^{2} \,n_{e} \,\,T_{e}^{3/2} }{9\,\,\pi
^{5/2}\,\,c^{3}\,\,T^{1/2}}\times\]
\begin{equation}
\times\left\{ \frac{2}{m_{ \bot} \,\,\tau _{ \bot} ^{(0)}} +
\frac{1}{m_{II} \,\,\tau _{II}^{(0)}}\right\} \,d\,%
 \underline {o}\, .
\end{equation}
Here, the fact that $(n_{k},T_{k})=(n_{e},T_{e})$ at $k=1$, 2, 3, and 4 was
taken into account.

From Eq.~(28), one can see that, if the concentration of electrons
and their temperatures in all the valleys are identical, the
dependence of the emission on the polarization disappears. This
conclusion is in agreement with the symmetry of the problem and
remains valid as far as the influence of the electric field on the
symmetry of the distribution function of electrons over the energy
can be neglected. (In this connection, see the consideration presented below.)

Now, let us come back to the case where the electric field is directed along
the unit vector $\vec{l}_{1}={\frac{{1}}{\sqrt{3}}}(1,1,1)$. Then, the
values of the parameters $n_{k}$ and $T_{k}$ are identical for valleys 2, 3,
and 4, but differ from the relevant magnitudes for $n_{1}$ and $T_{1}$. If
formula (\ref{eq23}) is written down in the form
\begin{equation}
\label{eq27} W^{( - )} = {\sum\limits_{k = 1}^{4} {W^{( - )}}}
(n_{k} ,T_{k} ,\varphi _{k} ),
\end{equation}
then, in the case concerned, we can expand it as follows:
\[ W^{( - )} = {\sum\limits_{k = 1}^{4} {W^{( - )}}} (n_{2} ,T_{2}
,\varphi _{k} ) +\]
\begin{equation}
\label{eq28} + W^{( - )}(n_{1} ,T_{1} ,\varphi _{1} )
 - W^{( - )}(n_{2} ,T_{2} ,\varphi _{1} ).
\end{equation}
That is, to the contributions made by three valleys ($k=2$, 3, and 4) with
identical parameters ${\left\{ {n_{k},T_{k}}\right\} }={\left\{ {n_{2},T_{2}}%
\right\} }$, we added and subtracted the contribution of the first
valley with the same parameters. Then, the sum in Eq.~(30)
coincides with
expression (28), where the substitution ${\left\{ {n_{k},T_{k}}%
\right\} }\leftrightarrow {\left\{ {n_{e},T_{e}}\right\} }$ is made, and
does not depend on the polarization.

The dependence of the emission on the polarization is governed
by two last terms in Eq.~(30):
\[
W^{( - )}(n_{1} ,T_{1} ,\varphi _{1}^{} )
 - W^{( - )}(n_{2} ,T_{2} ,\varphi _{1}^{} ) =
\]
\[
= \frac{4e_{0}^{2}}{3\pi ^{5/2}c^{3}T^{1/2}}\left( n_{1}
T_{1}^{3/2} - n_{2} T_{2}^{3/2} \right)\times\]
\begin{equation}
\label{eq29}\times {\left\{ {{\frac{{\sin ^{2}\varphi _{1}}} {{m_{
\bot} \tau _{ \bot }^{(0)}}} } + {\frac{{\cos ^{2}\varphi _{1}}}
{{m_{\parallel}  \tau _{\parallel} ^{(0)}}} }} \right\}}
d\underline {o}.
\end{equation}
Expressing $\sin^{2}\varphi_{1}$ and $\cos^{2}\varphi_{1}$ in terms of $%
\cos2\varphi_{1}$ and substituting expression (\ref{eq29}) into Eq.~(\ref%
{eq28}), we obtain
\[
W^{( - )} = W\left\{ {\,n_{2} ,T_{2}}  \right\} +
\frac{4\,e_{0}^{2} }{3\,\pi ^{5/2}\,c^{3}\,T^{1/2}}\times\]
\[\times(\,n_{1} T_{1}^{3/2}
- n_{2}^{} T_{2}^{3/2} )\left\{ \frac{1}{m_{ \bot}  \,\tau _{
\bot} ^{(0)}} + \frac{1}{m_{\parallel}  \,\tau _{\parallel}
^{(0)}} \right\}\,d\underline {o}  +\]
\[
 + \frac{4\,e_{0}^{2}} {3\,\pi ^{5/2}\,c^{3}\,T^{1/2}}\,(\,n_{1}
\,T_{1}^{3/2} - n_{2}^{} \,T_{2}^{3/2})\times\]
\begin{equation}
\label{eq30} \times\left\{\frac{1}{m_{\parallel} \,\tau
_{\parallel} ^{(0)}} - \frac{1}{m_{ \bot} \,\tau _{ \bot} ^{(0)}}
 \right\}\,\cos \;\;2\;\,\varphi _{1} \,  d\underline {o}.
\end{equation}
We are interested in the sign of the coefficient of $\cos 2\varphi _{1}$.
The matter is that, in the experiment with high-resistance specimens, the
growth of the electric field can be accompanied by the change of the sign of
this coefficient. The growth of the field increases the electron
temperature, and the growth of the latter can result in the change of the
scattering mechanism. Therefore, we want to elucidate whether and, if yes,
how the sign of this coefficient depends on the scattering mechanism. In
this section, we deal with acoustic scattering. For the corresponding
scenario,
\begin{equation}
\label{eq31} {\frac{{1}}{{m_{\parallel}  \tau _{\parallel}
^{(0)}}} }
 - {\frac{{1}}{{m_{ \bot}  \tau _{ \bot} ^{(0)}}} } \approx
 - {\frac{{1}}{{m_{ \bot}  \tau _{ \bot} ^{(0)}}} } <
 0.
\end{equation}
Therefore, the sign of the coefficient of $\cos 2\varphi _{1}$ is
determined by the sign of the expression $n_{1}T_{1}^{{3/2}%
}-n_{2}T_{2}^{3/2} $. Since valley 1 is \textquotedblleft
cold\textquotedblright\ and valleys 2, 3, and 4 are \textquotedblleft
hot\textquotedblright , we have $T_{2}>T_{1}$. The question concerning
the relationship between the parameters $n_{1}$ and $n_{2}$ remains
unanswered.

The parameters $n_{k}$ and $T_{k}$ must be determined from the concentration
and energy balance equations, respectively. But, it should be bear in mind
that the very form of these equations is based on the assumption about the
Maxwellian distribution function of hot electrons in every valley. This
assumption, in its turn, admits that the electron-electron scattering
dominates over the electron-phonon one. The intensity of $e-e$ scattering is
known to be proportional to the squared electron concentration, while the
corresponding cross-section is reciprocal to the fourth degree of relative
electron-electron speed. Therefore, the situation can be realized in
high-resistance specimens (i.e. with a low concentration of electrons),
when the $e-e$ interaction \textquotedblleft Maxwellizes\textquotedblright\ the
distribution function in the range of medium energies ($\varepsilon _{\vec{p}%
}\sim T_{e}$), but plays an auxiliary role in the formation of the shape of
the high-energy \textquotedblleft tail\textquotedblright\ of the
distribution function \cite{5}. At the same time, since the intervalley
repopulation of electrons (from hotter valleys to colder ones) is governed
just by the \textquotedblleft tail\textquotedblright\ of the distribution
function, this repopulation in high-resistance specimens can be neglected,
so that the electron concentrations in all valleys can be considered
approximately equal to one another. In this case, taking inequalities (\ref%
{eq31}) and $T_{2}>T_{1}$ into account, we see that the sign of the
coefficient of $\cos 2\varphi _{1}$ is positive. We will analyze this
issue in more details after having considered, in the next section, the
Coulomb mechanism of scattering.

\section{Spontaneous Emission by Hot Electrons at Impurity
(Coulomb) Scattering}

In this section, we consider optical transitions in a system of free
electrons, provided that they are scattered by charged impurities (ions).
The electron--ion interaction potential looks like
\begin{equation}
\label{eq32} U(\vec {r}) = \frac{e_{0}^{2}} {\chi _{0} r}\exp ( -
r/ r_{\rm D}),
\end{equation}
where $r_{\rm D}$ is the Debye radius. The expression for the
integral of electron--ion collision in the presence of an
electromagnetic wave and provided the dispersion law (\ref{eq1})
was derived by us in work \cite{2}. For electrons from the $i$-th
valley, the collision integral can be written down in the form
\[
 \hat {I}f^{(i)} = {\frac{{4e_{0}^{4}}} {{\chi _{0}^{2}}} }N_{\rm D}
{\sum\limits_{l = - \infty} ^{\infty}  {\int {d\vec
{p}{\frac{{f^{(i)}({\vec {p}}\,') - f^{(i)}(\vec
{p}\,)}}{{{\left\{ {(\vec {p} - {\vec {p}}')^{2} + ({{\hbar}
\mathord{\left/ {\vphantom {{\hbar}  {r_{\rm D}}} } \right.
\kern-\nulldelimiterspace} {r_{\rm D}}} )^{2}{\left. {} \right\}}
^{2}} \right.}}}}}}}  \times \]
\begin{equation}
\label{eq33} \times J_{l}^{2} \left( {{\frac{{e_{0}}} {{c\hbar
\omega }}}{\sum\limits_{\alpha = 1}^{3} {A_{\alpha} ^{(0)}
{\frac{{p_{\alpha}  - {p}'_{\alpha}} } {{m_{\alpha}} } }}}}
\right)\delta (\varepsilon _{\vec {p}} - \varepsilon _{{\vec
{p}}'} - l\hbar \omega ) ,
\end{equation}
where $N_{\rm D}$ is the concentration of ionized impurities.
Using the collision integral (\ref{eq33}) and applying the same
approximation as that in the derivation of Eq.~(\ref{eq11}), we
obtain
\[ P = \pm \hbar \omega {\frac{{4e_{0}^{4} N{}_{\rm D}}}{{\chi
_{0}^{2}}} }\int {d\vec {p}\,f^{(i)}(\vec {p}\,)} \times\]
\begin{equation}
\label{eq34} \times\int {d{\vec {p}}\,'{\frac{{\left(
{{\frac{{e_{0} \gamma}} {{2m_{ \bot}  \omega \,c}}}}
\right)^{2}\delta {\left\{ {\varepsilon _{\vec {p}} - \varepsilon
_{{\vec {p}}\,'} \pm \hbar \omega} \right\}}}}{{{\left\{ {\,(\vec
{p} - {\vec {p}}\,')^{2} + \left( {{{\hbar}  \mathord{\left/
{\vphantom {{\hbar}  {r_{\rm D}}} } \right.
\kern-\nulldelimiterspace} {r_{\rm D}}} } \right)^{2}}
\right\}}^{2}}}}^{}}.
\end{equation}
Impurity scattering is elastic. Therefore, $P_{0}=0$ (see Eq.~(\ref{eq10})),
and $\Delta P_{i}$ coincides with $P_{i}$. The parameter $\gamma $ in Eq.~(%
\ref{eq34}) stands for the quantity given by formula (\ref{eq5}) which
concerns the $i$-th valley.

After substituting formula (\ref{eq12}) into Eq.~(\ref{eq34}) and passing
to the deformed coordinate system, where the electron energy is independent
of angles, all the integrals over the remaining angles can be calculated
(see work \cite{2}), and the following result is obtained:
\[P_{i} ( + ) = \frac{e_{0}^{6} N_{\rm D} n_{i}} {4\chi _{0}^{2}
\,c^{2}\hbar \omega}\left( \frac{2\pi \,m_{\parallel}} {T_{i}}
\right)^{1/2}\frac{{A^{(0)}}^{2}}{(m_{\parallel} - m_{ \bot}
)^{2}}\times\]
\begin{equation}
\label{eq35} \times\int\limits_{0}^{\infty}  \frac{dx\,e^{ - x}\{
\Psi (q_{\max} ^{ +}  ) + \Psi (q_{\min} ^{ +}  ) \}}{\sqrt {x(x +
\hbar \omega / T_{i}  )}},
\end{equation}
Here, the notations
\[
 \Psi (q) = B_{1} (q) + (\vec {i}_{0} \,\vec {g}_{0} )^{2}{\left[ { - B_{1}
(q) + {\frac{{2m_{ \bot}} } {{m_{\vert \,\vert \,}}} }B_{2} (q)}
\right]},
\]
\[ q_{\max} ^{ +}  = {\frac{{(2m_{ \bot}  T_{i} )^{{{1}
\mathord{\left/ {\vphantom {{1} {2}}} \right.
\kern-\nulldelimiterspace} {2}}}}}{{\hbar }}}{\left[ {x^{{{1}
\mathord{\left/ {\vphantom {{1} {2}}} \right.
\kern-\nulldelimiterspace} {2}}} + (x + \hbar \omega / T_{i}
)^{{{1} \mathord{\left/ {\vphantom {{1} {2}}} \right.
\kern-\nulldelimiterspace} {2}}}} \right]},
\]
\begin{equation}
\label{eq36}
 q_{\min} ^{ +}  = {\frac{{(2m_{ \bot}  T_{i} )^{{{1} \mathord{\left/
{\vphantom {{1} {}}} \right. \kern-\nulldelimiterspace}
{}}2}}}{{\hbar }}}{\left[ { - x^{{{1} \mathord{\left/ {\vphantom
{{1} {2}}} \right. \kern-\nulldelimiterspace} {2}}} + (x + \hbar
\omega / T_{i} )^{{{1} \mathord{\left/ {\vphantom {{1} {2}}}
\right. \kern-\nulldelimiterspace} {2}}}} \right]}\,,
\end{equation}
\[
B_{1} = {\frac{{1}}{{b^{2}}}} + {\frac{{1 - b^{2}}}{{b^{3}}}}{\rm
arctg}\,\frac{1}{b},
\]
\[
B_{2} = - {\frac{{1}}{{1 + b^{2}}}} + {\frac{{1}}{{b^{}}}}{\rm
arctg}\,\frac{1}{b},
\]
\[
b^{2} = {\frac{{m_{ \bot}} } {{m_{\parallel}  - m_{ \bot} ^{}}}
}\left(1 + \frac{1}{(q\,r_{\rm D} )^{2}}\right)
\]
were introduced. The quantities $P_{i}(-)$ and $P_{i}(+)$ are
interconnected by an equation which is similar to Eq.~(13).

In the general case, expression (\ref{eq35}) is still cumbersome. The matter
is that, in the classical case ($\hbar \omega \ll T_{i}$), one has to take
the charge screening into account (otherwise, integral (\ref{eq35}) diverges),
whereas such a procedure is not necessary in the quantum-mechanical case.
Both cases include integral (\ref{eq35}); and, in both limit cases, integral (%
\ref{eq35}) is calculable to the end.

In particular, let us consider the classical case for the impurity scattering. By using formulas (\ref{eq35}) and (%
\ref{eq16}), we obtain the coefficient of absorption by free carriers
in the following form:
\begin{equation}
\label{eq37} K = {\frac{{3\pi ^{{{3} \mathord{\left/ {\vphantom
{{3} {2}}} \right. \kern-\nulldelimiterspace}
{2}}}}}{{2}}}{\frac{{e_{0}^{2}}} {{\chi _{0}^{{{1} \mathord{\left/
{\vphantom {{1} {2}}} \right. \kern-\nulldelimiterspace} {2}}}}}
}{\frac{{1}}{{c\omega ^{2}}}}{\sum\limits_{i - 1}^{4} {n_{i}}}
 {\left\{ {{\frac{{\sin ^{2}\varphi _{i}}} {{m_{ \bot} \,\tau
_{ \bot }(T_i)}} } + {\frac{{\cos ^{2}\varphi _{i}}}
{{m_{\parallel} \,\tau _{\parallel} (T_i)}} }} \right\}}.
\end{equation}
Here, $\tau_{\bot}(T_{i})$ and $\tau_{\parallel}(T_{i})$ are the
\textquotedblleft transverse\textquotedblright\ and
\textquotedblleft longitudinal\textquotedblright, respectively,
components of the relaxation time at the impurity scattering:
\[
{\frac{{1}}{{_{} \tau _{ \bot}  (T_{i} )}}} =
{\frac{{8}}{{3}}}\,{\frac{{e_{0}^{4} (2m_{\parallel}  )^{{{1}
\mathord{\left/ {\vphantom {{1} {2}}} \right.
\kern-\nulldelimiterspace} {2}}}}}{{\chi _{0}^{2} m_{ \bot}
T_{i}^{{{3} \mathord{\left/ {\vphantom {{3} {2}}} \right.
\kern-\nulldelimiterspace} {2}}}}} }\times\] \[N_{\rm D}
{\frac{{b_{0} }}{{2}}}{\left[ {b_{0} + (1 - b_{0}^2) \,{\rm
arctg}{\frac{{1}}{{b_{0}}} }} \right]}\ln \left( {C_{1} x_{\min}}
\right)^{ - 1} ,
\]
\[ \frac{1}{\tau _{\parallel}  (T_{i} )} =
\frac{8}{3}\,\frac{e_{0}^{4} (2m_{\parallel}  )^{ 1/2}}{\chi
_{0}^{2} m_{\parallel} T_{i}^{3/2}} \times\]
\begin{equation}
\label{eq38} \times N_{\rm D} {{{b_{0}}} {{}}}{\left[ { - b_{0} +
(1 + b_{0}^{2}) \,{\rm arctg}{\frac{{1}}{{b_{0}}} }} \right]}\ln
\left( {C_{1} x_{\min}} \right)^{ - 1}.
\end{equation}
Here, $b_{0}^{2}={m_{\bot}/(m_{\parallel}-m_{\bot})}$, $\ln
C_{1}=0.577\ldots$ is the Euler constant, and
\[
x_{\min}  = {\frac{{1}}{{8\,}}}{\frac{{\hbar ^{2}}}{{m_{ \bot}
T_{i} r_{\rm D} }}}.
\]
Unfortunately, in work \cite{2}, the expressions for
$\tau_{\bot}(T_{i})$ and $\tau_{\parallel}(T_{i})$ were confused.

In the same classical case within the same calculation scheme as in
the derivation of formula (\ref{eq23}), we obtain that the energy of
spontaneous emission by hot electrons on the impurity (Coulomb)
scattering is as follows:
\begin{equation}
\label{eq39} W^{( - )} \!=\! \frac{3\,e_{0}^{2}}{16\,\pi
^{3/2}c^3}\sum\limits_{i = 1}^{4} {n_{i} } T_{i}\left\{ \frac{\sin
^{2}\varphi _{i} }{m_{ \bot}  \,\tau _{ \bot}  (T_{i} )} \!+\!
\frac{\cos ^{2}\varphi _{i}}{m_{\parallel} \,\tau _{\parallel}
(T)_{i}} \right\}\! d\underline {o}  .
\end{equation}

If the electron-heating electric field is directed along the unit vector $%
\vec{l}_{1}={\frac{{1}}{\sqrt{3}}}(1,1,1)$, the dependence of the
emission on the polarization, similarly to Eq.~(\ref{eq29}) in the case of
the impurity scattering, reads
\[
 W^{( - )}(n_{1} ,T_{1}^{} ,\varphi _{1} ) - W^{( - )}(n_{2} ,T_{2} ,\varphi
_{1} ) =
\]
\[
 = \frac{3\,e_{0}^{2}}{16\,\pi ^{3/2}c^3}\Biggl\{
\left[\frac{n_{1}T_{1}} {m_{ \bot}  \,\tau _{ \bot}  (T_{1} )} -
\frac{n_{2} T_{2}} {m_{\, \bot}  \,\tau _{ \bot} (T_{2}
)}\right]\times\]
\begin{equation}
\label{eq40} \times\sin ^{2}\varphi _{1} + \left[\frac{n_{1}
\,T_{1}} {m_{\parallel} \tau _{\parallel}  (T_{1} )} - \frac{n_{2}
\,T_{2}} {m_{\parallel}  \tau _{\parallel} (T_{2} )}\right]\,\cos
^{2}\varphi _{1} \Biggr\}\,d\underline {o}.
\end{equation}

Let the repopulation of valleys be
inessential ($n_{1}\approx n_{2}$) owing to the reasons discussed above. Then, having expressed $\sin ^{2}\varphi
_{1}$ and $\cos ^{2}\varphi _{1}$ in terms of $\cos 2\varphi _{1}$ and
taking into account that -- according to Eq.~(\ref{eq38}) -- $m_{\bot }\tau
_{\bot }\ll m_{\parallel }\tau _{\parallel }$, we see that the coefficient of $%
\cos 2\varphi _{1}$ is negative. This means that the change of the scattering
mechanism along with the temperature variation (e.g., a transition from
the impurity scattering to the acoustic one) can induce the change of the sign of the coefficient
of $\cos 2\varphi _{1}$. Different signs of this coefficient at
the impurity and acoustic scatterings are associated with the fact that the
impurity scattering intensity falls down, while that of the acoustic
scattering grows up with increase in the electron temperature.

\section{Polarization Dependences Connected With the Distortion of the
Electron Distribution Function}

As the experiment shows, if the fields applied to $n$-Ge are strong, then,
even provided that the electron-heating electric field is oriented in the
symmetric -- with respect to the valleys -- direction $\vec{l}_{0}={\frac{{1}%
}{\sqrt{3}}}(1,0.0)$, the polarization dependence of the emission by hot
electrons is observed (see a more detailed discussion below). In this case,
the origin of symmetry violation can be only the electric field itself.
Since neither the scattering anisotropy nor the anisotropy of the dispersion
law play any important role in this situation, we consider the
influence of the electric field on the polarization dependence of the
emission by hot electrons in the framework of a simpler model. Namely, we
consider a mono-valley model with the dispersion law
\begin{equation}
\label{eq41} \varepsilon _{p}^{} = p^{2} / 2m
\end{equation}
and the isotropic acoustic scattering, where -- instead of
Eq.~(\ref{eq9}) -- the equation
\begin{equation}
\label{eq42} W_{a} = {\frac{{T}}{{4\pi ^{2}\hbar ^{4}\rho
\,s_{\vert \,\vert} ^{2} }}}{\Sigma_{d} {}}  = {\rm const}
\end{equation}
is now valid. In this model, if a strong electric field is applied, the
expression for the electron distribution function $f(\vec{p})$ should be
searched in the form
\begin{equation}\label{45}
 f(\vec {p}\,) = f_{0} (\varepsilon _{p} ) + f_{1} (\varepsilon _{p} ) \mathcal{P}_{1}
(\cos\theta )  + f_{2} (\varepsilon _{p} )\mathcal{P}_{2}
(\cos\theta  ) + \ldots ,
\end{equation}
where $P_{n}(\cos \theta )$ are Legendre polynomials, and $\theta
$ is the angle between the field $\vec{F}$ and the momentum
$\vec{p}$ direction. It is worth noting that, in the general case
-- i.e. if the dispersion law (1) and the scattering mechanism
(\ref{eq9}) were relevant, -- we would
have, instead of expression (\ref{45}), the expansion%
\begin{equation}
\label{eq43} f(\vec {p}\,) = {\sum\limits_{lm} {f_{lm}}}
(\varepsilon _{p} )\Upsilon _{lm} (\theta ,\varphi ),
\end{equation}
where $\Upsilon _{lm}(\theta ,\varphi )$ are spherical functions.
However, such a generalization makes the calculations excessively
cumbersome but does not change, in fact, the result of our
estimation of the influence of the distribution function
distortion by the field on the polarization dependences of
the emission by hot electrons. In the single-quantum approximation, the energy absorbed and emitted by electrons
in a unit time interval
is given by expression (\ref{eq11}), which is valid
for an arbitrary distribution function.

Earlier, we adopted the Maxwell function (\ref{eq12}) as the
distribution function with the dispersion law (1). In so doing, we
did not consider the influence of the field $\vec{F}$ on the
distribution function symmetry (although the field did affect the
electron temperature). In our present model, the violation of the
distribution function symmetry in expansion (45) is characterized
by the expressions $f_{1}(\varepsilon _{p})\mathcal{P}_{1}(\cos
\theta )$, $f_{2}(\varepsilon _{p})\mathcal{P}_{2}(\cos \theta )$,
and so on. The term proportional to $f_{1}(\varepsilon _{p})$,
owing to its oddness, gives no contribution to expression
(\ref{eq11}). We should note that expansion (45) is actually an
expansion in the dimensionless parameter $e_{0}F\,\tau /p$, i.e.
in the ratio between the field-induced momentum gain of the
electron within the mean free time and the electron momentum itself \cite{5}%
. Therefore, for $f_{2}(\varepsilon _{p})$, according to the results of work
\cite{5}, we obtain
\begin{equation}
\label{eq44} f_{2} (\varepsilon _{\rho}  ) \approx
{\frac{{2}}{{3}}}\left( {e_{0} F} \right)^{2}\tau (\rho
)p{\frac{{d}}{{d\rho}} }\left( {{\frac{{\tau (\rho
)}}{{p}}}{\frac{{df_{0}}} {{d\rho}} }} \right),
\end{equation}
where$\mathrm{\ }$the notation
\[
{\frac{{1}}{{\tau (\rho )}}} = {\frac{{1}}{{\pi}}
}\,{\frac{{mT}}{{\rho \,s_{\parallel} ^{2} \hbar ^{4}}}}p \equiv
{\frac{{1}}{{\tau ^{(0)}}}}\left( {{\frac{{\varepsilon}} {{T}}}}
\right)^{{{1} \mathord{\left/ {\vphantom {{1} {2}}} \right.
\kern-\nulldelimiterspace} {2}}}.
\]
was introduced.

Substituting expansion (45) into Eq.~(\ref{eq11}) and integrating
over the angles, we obtain
\[
\Delta P( - ) = \frac{8\pi ^{2}}{3}\frac{e_{0}^{2}
(A^{(0)})^{2}W_{a}}{m\hbar \omega \,c^{2}}\times\]
\[\times \int\limits_{\sqrt {2m\hbar \omega}}^{\infty} {d\rho
\,p^{2}} \sqrt {p^{2} - 2m\hbar \omega} \Biggr\{ \,(p^{2}  -
2m\hbar \omega )f_{0} (\varepsilon _{\rho } ) +
\]
\[+\frac{1}{5}\mathcal{P}_{2}(\cos\theta _{0})
\frac{(eF\tau ^{(0)})^{2}}{mT_{e}^{2}}T\,\Biggl[  - a^{3}e^{a}
\frac{d}{da}\left( \frac{K_{1} (a)}{a} \right) +
\]
\begin{equation}
+\frac{1}{2}(1 + 4a)ae^{a}K_{1}(a) \Biggr]\Biggr\} _{a = {\hbar
\omega}/ {2T_{e}}}.
\end{equation}
Here, the quantities $\Delta P(-)$ and $\Delta P(+)$ are connected
by relationship (13), and $\theta_{0}$ is the angle between the
polarization unit vector $\vec{g}_{0}$ and the field $\vec{F}$.

Now, in Eq.~(48), we substitute $f_{0}(\varepsilon _{\rho })$ by
the
Maxwell function with the effective electron temperature $T_{e}$ and $%
f_{2}(\varepsilon _{\rho })$ by expression (\ref{eq44}). After the integration,
we have
\[
\Delta P( - ) = \frac{2}{3\sqrt {\pi}} \frac{e_{0}^{2}
(A^{(0)})^{2}n}{m\,\tau ^{(0)}\hbar \,\omega
\,c^{2}}\frac{T_{e}^{3/2}}{T^{1/2}}\times\]
\[\times e^{ - \hbar
\omega /T_{e} }\Biggr\{ { - a^{3}e^{a}} \frac{d}{da}\left(
\frac{K_{1} (a)}{a} \right) + \frac{2}{15} \mathcal{P}(\cos\theta
_{0} )\times\]
\[  \times \frac{(eF\tau ^{(0)})^{2}}{mT_{e}^{2}}
T\Biggr[  - a^{3}e^{a}\frac{d}{da}\left( \frac{K_{1}
(a)}{a}\right) +\]
\begin{equation}
+ \frac{1}{2}(1 + 4a)ae^{a}K_{1} (a) \Biggr]\Biggr\} _{a = \hbar
\omega / {2T_{e}}}. \label{49}
\end{equation}
It is of interest to compare the first term in Eq.~(49) with
expression (13). Putting $m_{\bot }\tau _{\bot }^{(0)}=m_{\Vert
}\tau _{\Vert }^{(0)}$, $T_{i}=T_{e}$, and $n_{i}=n$, we get a
complete
coincidence. Formula (13) determines the contribution of a single $i$%
-th valley to the absorption and emission processes.
Much more interesting for a comparison with expression (13) is not
to reduce the model to the isotropic case but to take
the identical contributions from all valleys into consideration, which
occurs if the field $\vec{F}$ is oriented along the (1,0,0)
direction. In this case, the axes $T_{i}$ are equal to each other
(and to $T_{e}$), and all $n_{i}$ are equal to each other (and to
$n/4$, where $n$ is the total concentration over all valleys).
Doing such a comparison, we see that the first term in Eq.~(49)
coincides with the expression for an anisotropic multivalley case,
provided the formal substitution
\begin{equation}
\label{eq45} {\frac{{1}}{{m\tau ^{(0)}}}} \to
{\frac{{1}}{{3}}}\left( {{\frac{{2}}{{m_{ \bot}  \tau _{ \bot}
^{(0)}}} } + {\frac{{1}}{{m_{\parallel}  \tau _{\parallel}
^{(0)}}} }} \right).
\end{equation}
The matter is about the orientation of the heating field in the symmetric
direction.

Knowing expressions for the quantities $P(\pm )$ in the model,
where the function $f(\vec{p})$ is given by expansion (45), it is
easy to find the energy of spontaneous emission by hot
electrons. Similarly to the procedure of deriving Eq.~(23) from
Eq.~(13), Eq.~(49) yields
\[
W^{( - )} = \frac{2e_{0} \,nT_{e}^{3/2}}{3\pi ^{5/2}c^{2}T^{1/2}}
\frac{1}{m\tau ^{(0)}}\times\]
\[\times\Biggl\{ - a^{3}e^{ -
a}\frac{d}{da}\left(\frac{K_{1} (a)}{a} \right) +
\frac{2}{15}\mathcal{P}_{2} (\cos\theta _{0} )\times
\]
\[\times
\frac{(e_{0} F\tau ^{(0)})^{2}}{mT_{e}^{2}}T \Biggr[- a^{3}e^{ -
a}\frac{d}{da}(\frac{K_{1} (a)}{a}) +\]
\begin{equation}
\label{eq46}
 \frac{1 + 4a}{2}ae^{ - a}K_{1}(a)\Biggr]\Biggr\} _{a = {\hbar\omega}
 /{2T_{e}}}
d\underline {o}.
\end{equation}

In the multivalley case, provided that the field $\vec{F}$ is
oriented in the direction (1,0,0), expression (\ref{eq22}) is
transformed -- by formally
applying substitution (\ref{eq45}) -- into the first summand in Eq.~(\ref%
{eq46}). This summand does not contain the polarization
dependence. Such a dependence appears, as is evident, if the
dependence $f_{2}(\varepsilon _{\rho })$ is made allowance for in
Eq.~(45), i.e. if going beyond the frame of the diffusion
approximation. In other words, the polarization dependence of
the emission by hot electrons, provided that the field is
oriented in the symmetric direction, is associated with the
symmetry violation of the electron distribution function over the
energy (the even part of the distribution function!).

In the classical case (if $a\ll 1$), expression (\ref{eq46})
becomes simpler:
\[
W^{( - )} = \frac{4e_{0}^{2} \,nT_{e}^{3/2}} {3\pi
^{5/2}c^{2}T^{1/2}}\frac{1}{m\tau ^{(0)}}\times\]
\begin{equation}
\times\left\{ {1} + \frac{(e_{0} F\tau ^{(0)})^{2}}{6mT_{e}^{2}} T
\mathcal{P}_{2} (\cos\theta _{0} )\right\} \,d\underline {o} .
\label{52}
\end{equation}
By the order of magnitude, the second term in the braces is equal to the
squared ratio between the drift and thermal speeds.

\section{Discussion of Results. Comparison with Experimental Data}

We have demonstrated in the previous sections that, in the $n$-Ge case, the
change of the sign is possible for the coefficient that characterizes the
angular dependence of the polarization, if the electron-heading field is
oriented along the (1,1,1)-axis. This change is caused by the change of the
scattering mechanism, which, in its turn, is caused by the increase of the
electron temperature with the growing field. Figure~1 illustrates such a
sign change observed experimentally. The measurement technique was described
in work \cite{3}. The curves in this figure correspond to specimens with a
charge carrier concentration of $2.5\times10^{15}$~\textrm{cm}$^{-3}$.

\begin{figure}[t]
\centering\includegraphics[width=0.9\columnwidth]{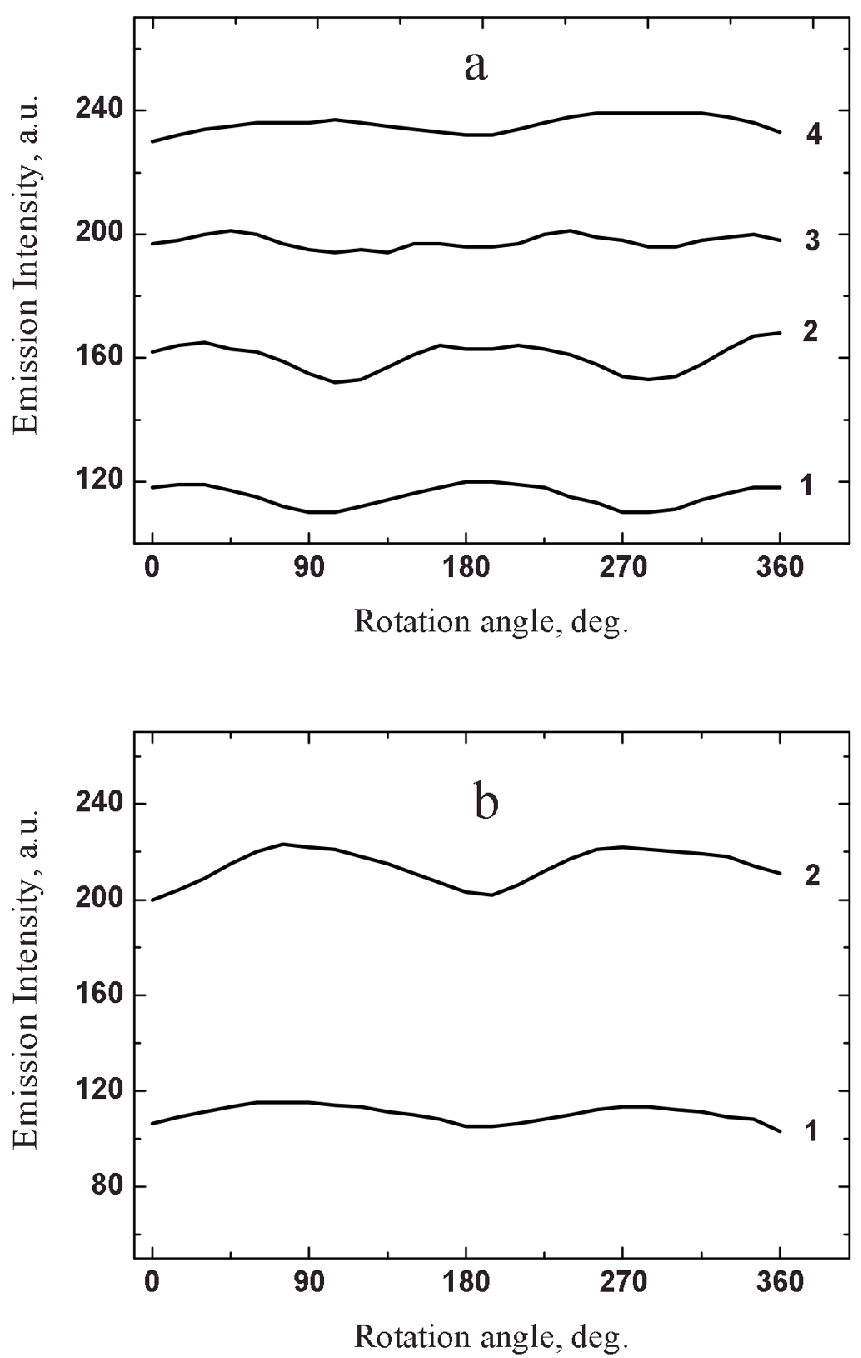}
\caption{\label{fig1}Polarization angular
dependences of the emission by hot charge carriers in an $n$-Ge
specimen (the crystallographic direction $\langle
111\rangle$ and the carrier concentration $n\approx2.5\times10^{15}$~\textrm{%
cm}$^{-3}$) for various values of the heating electric field: (panel
$a$) 10 (\textit{1}), 20 (\textit{2}), 40 (\textit{3}), and 75~V/cm
(\textit{4});
(panel $b)$ 80 (\textit{1}) and 140~V/cm (\textit{2})}
\end{figure}

Figure~1 depicts the dependences of the emission intensity by hot
electrons on the polarization rotation angle at various values of the
heating field (10, 20, 80, and 140~V). One can see that the growth of the
field gives rise to the modification of the curve character. In particular,
the angles, which correspond to observable minima at weak fields, become
characterized by the maximal intensity at higher ones.

It is worth noting that the mechanism of scattering can change due
to the variation of the lattice temperature as well. In this case,
the change of the sign of the coefficient mentioned above is also
possible. This opportunity is illustrated in Fig.~2. In this figure,
exhibited are the angular dependences of the emission at various
temperatures. To illustrate the angular dependences, the curves are
somewhat shifted vertically with respect to one another, although
the intensities of the emission are approximately identical at
different temperatures.
The curves correspond to specimens with a carrier concentration of $%
2\times10^{14}$ \textrm{cm}$^{-3}$. The character of modifications
of the angular dependences is the same as in Fig.~1. An
opportunity for the angular dependence of the emission to
change its profile with the modification of the scattering
mechanism was pointed out in work~\cite{3}.

\begin{figure}[t]
\centering\includegraphics[width=0.9\columnwidth]{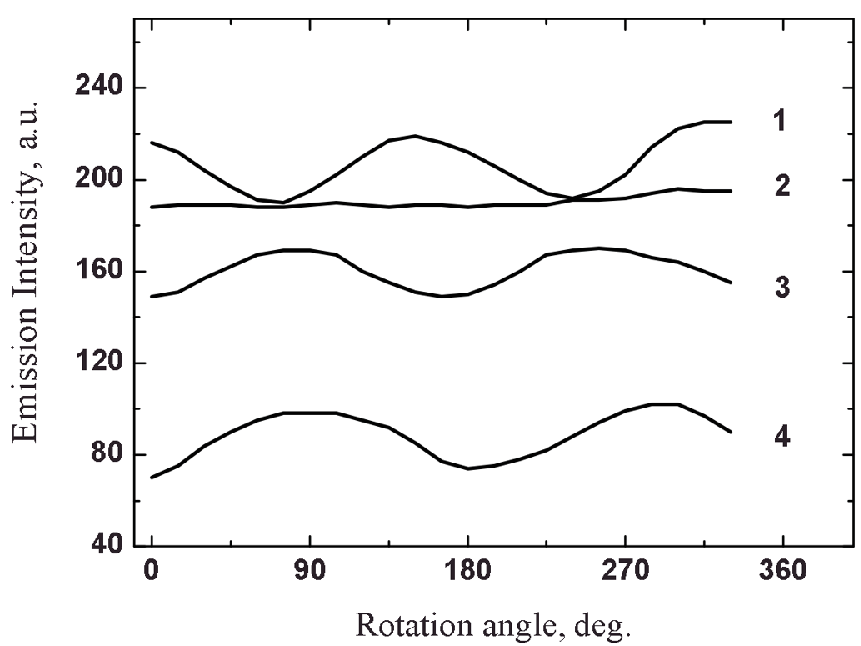}
\caption{\label{fig1}Polarization angular
dependences of the emission by hot charge carriers in an $n$-Ge
specimen (the crystallographic direction $\langle 111\rangle$,
$n\approx6\times10^{14}$~\textrm{cm}$^{-3}$, the applied
electric field $E=140~\mathrm{V/cm}$) for various temperatures: 6.6 (\textit{%
1}), 7.7 (\textit{2}), 14 (\textit{3}), and 76$~\mathrm{K}$ (\textit{4})}
\end{figure}
\begin{figure}[!]
\centering\includegraphics[width=0.9\columnwidth]{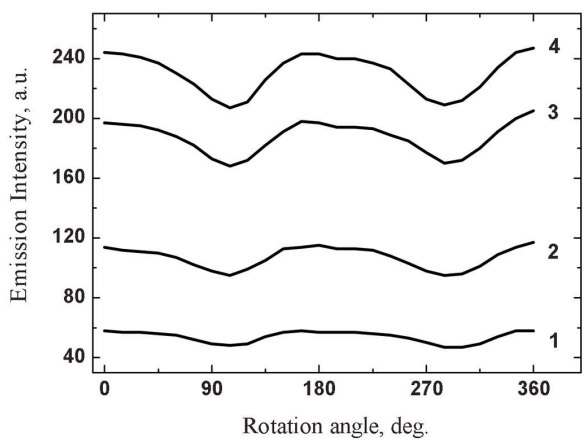}
\caption{\label{fig1}Polarization angular
dependences of the emission by hot charge
carriers in an $n$-Ge specimen (the symmetric crystallographic direction $%
\langle100\rangle$, $n\approx1.5\times10^{14}$~\textrm{cm}$^{-3}$)
for
various values of the heating electric field: 7 (\textit{1}), 10 (\textit{2}%
), 13 (\textit{3}), and 15~V/cm (\textit{4})}
\end{figure}

Rather unexpected was the appearance of po\-la\-ri\-za\-tion
dependence for the heating field oriented along the (1,0,0)-axis.
Provided such a field orientation, all valleys are characterized by
identical values of the concentration and the temperature of
electrons. Figure~3 illustrates the corresponding polarization
dependence obtained experimentally. Specimens
with a charge carrier concentration of $1.5\times 10^{14}$~\textrm{cm}$%
^{-3}$ were used. The figure exhibits the angular dependence of the
emission at various heating fields applied in the symmetric direction. The
specimen temperature was 5$~\mathrm{K}$. The curves evidently reveal the
angular dependence of the emission, although all the parameters of
valleys (the temperature and the concentration of electrons) are identical.
Such an unusual behavior can be associated with going beyond the traditional
approach which is reduced to the so-called diffusion approximation.

Note that a combined influence on the character of the angular
dependences can take place at low temperatures and relatively strong
fields, namely, the change of scattering mechanisms and the
violation of the conditions of the diffusion approximation.

Thus, we showed that in the case when the heating field is oriented
in the symmetric direction the polarization dependence is caused by
the distortion of the energy distribution function of electrons
(going beyond the range of validity of the diffusion approximation).

\vskip3mm The authors express their gratitude to O.G.~Sarbey for
his permanent interest to this study and for fruitful discussions.

\selectlanguage{russian}

\rezume{%
 ПОЛЯРИЗАЦІЙНІ ~~ЕФЕКТИ~~ В~~ ВИПРОМІНЮВАННІ \\І ПОГЛИНАННІ СВІТЛА
ГАРЯЧИМИ ЕЛЕКТРОНАМИ \\У БАГАТОДОЛИННИХ НАПІВПРОВІДНИКАХ}{П.М.
Томчук, В.М. Бондар} {Теоретично і експериментально на прикладі
$n$-Ge досліджено кутові залежності спонтанного випромінювання
гарячих електронів в багатодолинних напівпровідниках. Показано, що
зміна механізму розсіяння, зумовлена ростом електронної
температури, може приводити до зміни характеру кутової залежності
розсіяння. Вперше експериментально спостережено кутову залежність
випромінювання у випадку прикладання гріючого електрони поля
вздовж осі симетрії кристала (для $n$-Ge це вісь (1,0,0,)).
Побудована теорія такої залежності. Показано, що кутова залежність
випромінювання у випадку, коли електрони всіх долин мають однакову
концентрацію і температуру, пов'язана з порушенням симетрії
функції розподілу електронів за енергіями (в теорії це означає
вихід за рамки традиційного дифузійного наближення).}

\end{document}